\documentclass[12pt]{article}
\pdfoutput=1
\usepackage{lutypaper,graphicx,xcolor}
\usepackage[normalem]{ulem}
\usepackage[font=small,labelfont=bf]{caption}
\numberwithin{equation}{section} 

\renewcommand{\vec}[1]{{\mathbf{#1}}}
\newcommand{\HD}[1]{{\color{brown} \bf{#1}}}

\begin{document}

\begin{titlepage}

\title{Blowing in the Dark Matter Wind}

\author{Hannah Day$^*$\ \ 
Da Liu$^{\dagger,\ddagger}$\ \ 
Markus A. Luty$^\ddagger$\ \ 
Yue Zhao$^\diamondsuit$}

\address{$^*$Department of Physics, University of Illinois Urbana-Champaign\\
Urbana, IL 61801}

\address{$^\dagger$PITT PACC, University of Pittsburgh\\
Pittsburgh, PA, USA}

\address{$^\ddagger$Center for Quantum Mathematics and Physics (QMAP)\\
University of California, Davis, CA 95616}

\address{$^\diamondsuit$Department of Physics and Astronomy\\
University of Utah, Salt Lake City, UT 94103}

\begin{abstract}
Interactions between dark matter and ordinary matter will transfer momentum,
and therefore give rise to a force on ordinary matter due to the dark matter
`wind.' We show that this force can be maximal in a realistic model
of dark matter, meaning that 
an order-1 fraction of the dark matter momentum incident
on a target of ordinary matter is reflected.
The model consists of light ($m_\phi \lsim \text{eV}$) scalar dark matter with
an effective interaction $\phi^2 \bar{\psi}\psi$, 
where $\psi$ is an electron or nucleon field.
If the coupling is repulsive and sufficiently strong, 
the field $\phi$ is excluded 
from ordinary matter, analogous to the Meissner effect for photons
in a superconductor.
We show that there is a large region of parameter space that is compatible with
existing constraints, 
where the force is large enough to be detected by existing force
probes, such as satellite tests of the
equivalence principle and torsion balance experiments.
However, shielding of the dark matter by ordinary matter prevents
existing experiments from being sensitive to the dark matter force.  
We show that precise measurements of spacecraft trajectories
proposed to test long distance modifications of gravity
are sensitive to this force for a wide range of parameters.
\end{abstract}

\end{titlepage}

\noindent
\section{Introduction}
Understanding the microscopic nature of dark matter is one of the major 
open problems in particle physics and cosmology.
Dark matter has been detected only through its gravitational interactions,
but additional interactions are needed in order to explain its relic abundance,
giving hope that we can find additional signals of dark matter in the laboratory
or in astrophysical or cosmological observations.
The allowed dark 
matter masses and interactions span a vast parameter space, 
and it is important to carry out experimental searches for as wide a range of models as possible.

In this paper, we consider a new probe of dark matter: the force on ordinary
matter due to an incident flux of dark matter---the dark matter `wind.'
Due to the motion of the solar system
in the galaxy relative to the dark 
matter halo, the local dark matter is moving relative
to the solar system with an average speed 
$v_\text{DM} \sim 300~\text{km}/\text{s} \sim 10^{-3}c$.
We can estimate the maximal force that could be exerted by the dark matter
wind by assuming that the full flux of dark matter on an object is
reflected by the object.
If the dark matter interacted strongly with a macroscopic target
with cross-sectional area $A$, the force exerted by the dark
matter on the target is given by
\[
\eql{Fmax}
F_\text{max} \sim \rho^{\vphantom 2}_\text{DM} v_\text{DM}^2 A,
\]
where $\rho_\text{DM} \sim 0.3$~GeV$/$cm$^3$ is the
dark matter density {near the earth. 
The corresponding acceleration of a  solid sphere of radius $R$ and density $\rho$
is
\[
\eql{amax}
a_\text{max} \sim 10^{-12}~\text{m}/\text{s}^2 \ggap
\left( \frac{R}{\text{cm}} \right)^{-1} 
\left( \frac{\rho}{10~\text{g}/\text{cm}^3} \right)^{-1}.
\]
These accelerations are large enough to be detected using sensitive
low-frequency force probes, such 
as those used to test the equivalence 
principle~\cite{Wagner:2012ui, Schlamminger:2007ht,Lee:2020zjt}.
On the other hand, this interaction also means
that any surrounding ordinary matter (including the earth's atmosphere)
can partially or completely shield the force probes from the dark matter.
We will see that this effect makes existing experiments insensitive
to the dark matter force.
We will comment  on the possibility of future experiments 
that can detect the effect.

The interactions of dark matter with ordinary matter are
strongly constrained by direct dark matter searches, as well as
cosmological and astrophysical constraints.
Nonetheless, we will show that there is a model of dark
matter where the force on ordinary matter can be maximal,
and which is compatible with all these constraints.
In this model, dark matter is described by a light scalar field $\phi$
with an effective coupling
\[
\eql{theint}
\scr{L}_\text{int} = -\frac{1}{2f_\psi} \phi^2 \bar{\psi} \psi,
\]
where $\psi$ is an electron or nucleon field.
The absence of a Yukawa coupling $\phi\bar\psi \psi$
can be explained by a $\mathbb{Z}_2$ symmetry under which
$\phi \mapsto -\phi$.
In a region containing matter, we have $\avg{\bar{\psi}\psi} = n_\psi$ 
in the non-relativistic limit, 
where $n_\psi$ is the number density of $\psi$ particles~\footnote{Note that individual particle will scatter with dark matter particles with suppression of $1/f_\psi^2$, resulting very small force effect  in the parameter space considered in this paper $f_\psi \gtrsim 10^8$ GeV (see Fig.~\ref{fig:constraints} and discussion there).}.
Therefore, this interaction gives an additional contribution to the 
dark matter mass inside matter:
\[
\eql{Dem2}
\De m_\phi^2 = \frac{n_\psi}{f_\psi}.
\]
We will consider the case $f_\psi > 0$, so that $\De m_\phi^2 > 0$.
Inside matter, the relation between the energy and momentum
of a $\phi$ particle is modified:
\[
E_\text{matt} = \sqrt{k_\text{matt}^2 + m_\phi^2 + \De m_\phi^2}.
\]
If $\De m_\phi^2$ is sufficiently large, then the propagation
of dark matter is suppressed inside matter.
For a $\phi$ particle in vacuum with momentum $k$ incident on a region
of ordinary matter, energy conservation gives
\[
E_\text{matt} = E_\text{vac} = \sqrt{k_\text{vac}^2 + m_\phi^2},
\]
and we see that $k_\text{matt}^2 < 0$ for  $\De m_\phi^2 > k_\text{vac}^2$.
If this is satisfied, then the matter region is classically forbidden,
and the propagation of the dark matter is exponentially
suppressed.
This is a coherent scattering effect that is important when 
the density of scatterers is large compared to the de Broglie 
wavelength of the dark matter.
It is analogous to the Meissner effect for photons in a 
superconductor, so we call this the `dark matter Meissner effect.'

The shielding of the field $\phi$ in this model
by ordinary matter 
was previously discussed in \cite{Hees:2018fpg,Banerjee:2022sqg}.
However, they did not discuss the force from the dark matter wind
that is the focus of the present work.
The dark matter force on planets in this model was considered in 
\cite{Fukuda:2018omk},
but there result included a large coherent enhancement factor
that we believe is not present.

We will show that there is a large region of the 
parameter space of this model where the dark matter Meissner effect 
takes place,  which is  compatible with existing 
laboratory and astrophysical/cosmological
constraints.
This opens the exciting possibility that it may be possible to
directly measure the force exerted by dark matter on ordinary matter.
We emphasize that here we are discussing the average force
due to the motion of the dark matter, rather than the oscillatory
force due to coherent $\phi$ oscillations with frequency $\om = m_\phi$,
which has been previously considered in the literature~\cite{Graham:2015ifn, Carney:2019cio}.
We expect that we can approximate the wind force as time-independent
if we average over time scales
longer than the coherence time of these oscillations, which is
of order
\[
t_\text{coherence} \sim \frac{1}{m_\phi v_\text{DM}^2}
\sim \text{sec} \left( \frac{m_\phi}{10^{-8}\eV} \right)^{-1}.
\]

The strong interaction between ordinary matter and dark matter
makes the detection of the dark matter wind force possible, 
but it also means that ordinary matter can act as a shield for the 
dark matter wind.
Because of this shielding effect, we are not able to identify
any existing force experiments that are sensitive to this force.
We will give estimates for the size of the force, and discuss 
some possible future experiments that may be sensitive to it.
Further work is needed to design an experiment that is
sensitive to this force.

This paper is organized as follows.
In \S\ref{sec:Meissner}, we discuss the physics of the dark matter
Meissner effect. In \S\ref{sec:constraints} we review existing constraints on the model.
In \S\ref{sec:experiments} we give estimates for the size of 
the force on potential experimental targets.
Our conclusions are given in \S\ref{sec:conclusion}.
Appendices give additional details of our calculations,
and discuss UV completions and fine tuning.

\section{The Dark Matter Meissner Effect\label{sec:Meissner}}
In this section we discuss the dark matter Meissner effect in more
detail.
We begin with some simple estimates for the
the parameter regime where the effect takes place and can lead
to a maximal force on a target.
We then explain the methods used to perform precise calculations
of the force, and give some parametric estimates for various
limiting cases.

\subsection{Estimates}
In the presence of matter, coherent scattering effects can be important
if the density of scatterers is sufficiently high.
For example, for the earth's atmosphere,  the ratio between the DM de Broglie wavelength and the distance among two atoms  is roughly given by
\[
\eql{dBlarge}
\frac{\bar{\la}_\text{dB}}{n_\text{atm}^{-1/3}}
\sim 10^5 \left( \frac{m_\phi}{\text{eV}} \right)^{-1},
\]
where $\bar{\la}_\text{dB} = 1/m_\phi v_\phi$ is the reduced
de Broglie wavelength of the dark matter particles in vacuum
and $n_\text{atm}$ is the number density of nucleons near the
earth's surface.
As long as this ratio is large,  we can approximate the matter as continuous.
Inside matter, we then have
\[
\avg{\bar\psi \psi} = n_\psi \sim \frac{\rho_\text{matter}}{m_p},
\]
where $n_\psi$ is the number density of $\psi$ particles 
(nucleons or electrons) in matter.%
\footnote{In ordinary matter,
the electron number density is roughly half of the nucleon number density.}
Therefore, as discussed in the introduction, dark matter will be excluded
from regions of sufficiently high density if 
$\De m_\phi^2 > k_\text{vac}^2$, where $\De m_\phi^2$ is the
matter contribution to the dark matter mass.
This condition is satisfied for
\[
\eql{DMMeisscond}
\frac{1}{f} \gsim 2 \times 10^{-8}~\text{GeV}^{-1}
\left( \frac{m_\phi}{\text{eV}} \right)^2
\left( \frac{\rho_\text{matter}}{10~\text{g}/\text{cm}^3} \right).
\]
This can be written more intuitively as
\[
\bar{\la}_\text{dB} \gsim L_\text{skin}.
\]

If this condition is satisfied, then the propagation of dark
matter into the target has an exponential suppression
$\sim e^{-d/L_\text{skin}}$, where $d$ is the distance into the
target and%
\footnote{The second equality in \Eq{Lskin} assumes that 
$\De m_\phi^2 \gsim k_\text{vac}^2$.}
\[
L_\text{skin} = \frac{1}{|k_\text{matt}|} \sim
\left( \frac{f}{n_\psi} \right)^{1/2}
\sim 3 \times 10^{-2}~\text{cm}
\left( \frac{1/f}{10^{-8}~\text{GeV}^{-1}} \right)^{-1/2}
\left( \frac{\rho_\text{matter}}{10~\text{g}/\text{cm}^3} \right)^{-1/2}.
\eql{Lskin}
\]
is the skin depth associated with the dark matter Meissner effect.
In order for the force to be maximal, 
for a target of linear size $R$
we also require
\[
\eql{skincond}
R \gsim L_\text{skin},
\]
so that the exponential suppression causes an order-1 fraction of
the dark matter to be reflected from the target.
If \Eqs{DMMeisscond} and \eq{skincond} are both satisfied, 
we expect the force to be maximal, so that the estimate \Eq{amax} holds.

Assuming we are interested in experimental targets with the density
of ordinary matter  ($\sim \text{g}/\text{cm}^3$)
and with size $\sim 1$--$10$~cm, 
the region of parameters that we can hope to probe is roughly
(see Fig.~\ref{fig:constraints})
\[
10^{-8}~\text{eV} \lsim m_\phi \lsim 10~\text{eV},
\qquad
10^{-13}~\text{GeV}^{-1} \lsim \frac{1}{f} \lsim 10^{-8}~\text{GeV}^{-1}.
\]
The upper bounds on $m_\phi$ and $1/f$ come from the requirement that
the force be maximal;
the lower bounds come from constraints on the model from nucleosynthesis
and supernova cooling.
These constraints will be reviewed in \S\ref{sec:constraints}.

\subsection{Quantitative Calculations}

We now discuss how to perform quantitative calculations of the
force due to the dark matter wind.
We consider a monochromatic wave of dark matter incident on a
target localized at the origin.
The force on the target can be computed 
using one of two approximations.
In the first, 
we approximate the dark matter as a classical field and 
find the classical scattering solution for the field 
in the presence of the target.
We can then compute the force on the target by computing the
momentum transferred to the target by the field.
The classical field approximation is valid for
$m_\phi \ll 10\eV$, which is satisfied in most (but not
all) of the phenomenologically interesting parameter space
of our model.
In the second approximation,
we treat the incident dark matter wave
as a superposition of $\phi$ particles, and compute the
scattering probability of these particles from the target
using non-relativistic quantum scattering theory.
We can then compute the force by adding up the momentum transferred
to the target by each scattering particle.
In Appendix A, we show that these two methods give identical results
in the non-relativistic limit.

We begin with the classical field picture.
We assume that far from the target the solution is a plane wave
in the $+z$ direction:
\[
\lim_{r \to \infty}
\phi(\vec{r}, t) = \Re\!\big[ \phi_0 e^{-i(\om t - k z)} \big],
\]
where
\[
\om = \sqrt{k^2 + m_\phi^2}
\]
is the frequency in vacuum.
To compute the force, we consider a steady state solution of the form
\[
\eql{steadyphi}
\phi(\vec{r}, t) = \Re\!\big[ e^{-i \om t} \psi(\vec{r}) \big].
\]
The equations of motion of the field in the presence of the target give
\[
\bigg( \grad^2 + k^2 - \frac{n_\psi(\vec{r})}{f} \bigg) 
\psi(\vec{r}) = 0,
\]
which we can rewrite as
\[
\eql{NRSE}
\bigg[ {-\frac{1}{2m_\phi}} \grad^2 + V_\text{eff}(\vec{r})
\bigg]
\psi(\vec{r}\gap) = E \psi(\vec{r}\gap),
\]
where
\[
\eql{Veff}
V_\text{eff}(\vec{r}) = \frac{n_\psi(\vec{r})}{2m_\phi f},
\qquad
E = \frac{k^2}{2m_\phi}.
\]
This is the time-independent non-relativistic Schr\"odinger equation 
that describes scattering solutions for a 
$\phi$ particle with potential $V_\text{eff}(\vec{r})$.
We are interested in the non-relativistic case ($k \ll m_\phi$), 
but it is interesting
that we obtain the non-relativistic Schr\"odinger
equation even if the system is relativistic.
More importantly for us, this allows us to use standard 
results from quantum mechanical scattering theory in the classical 
calculation, and will be used to show that the quantum
and classical calculations predict the same time-averaged force.

We illustrate this by consider a potential with spherical symmetry.
We expand
the wavefunction in 
spherical Bessel functions and Legendre polynomials (see Appendix A):
\[
\psi(\vec{r}) = \phi_0 \sum_\ell
i^\ell (2\ell + 1) \biggl[  j_\ell(kr)
+ \frac{S_\ell -1}{2} \bigl( j_\ell(kr) + i y_\ell(kr) \bigr) \biggr] P_\ell(\cos\th).
\]
Here $S_\ell$ is the quantum-mechanical $S$-matrix
element for the partial wave with angular momentum $\ell$, given by
\[
\eql{Sell}
S_\ell = e^{2i\de_\ell},
\]
where $\de_\ell$ is the phase shift for the partial wave $\ell$
in the effective potential \Eq{Veff}.
The result is that the time-averaged force on a target is given by
\[
\eql{Fzclassical}
F^z_\text{target} = -\frac{\pi \rho_\phi}{m_\phi^2} 
\sum_{\ell \, = \, 0}^\infty (\ell + 1)
\Big[ S_\ell^* S\sub{\ell + 1} + S_{\ell+1}^* S\sub{\ell} - 2 \Big],
\]
where 
\[
\eql{phinorm}
\rho_\phi = \sfrac 12 |\phi_0|^2 m_\phi^2
\]
is the density of dark matter.
Note that the force vanishes in the no-scattering limit 
$S_\ell \to 1$, as it must.
The derivation of this result is given in Appendix A
for both the classical and quantum pictures.
Formulas for $S_\ell$ for various geometries are given in
the appendix: see \Eq{Aelldefn} together with
\Eq{Alsolid} (for a solid sphere),
\Eq{Alhollow} (for a hollow sphere).
%

We now use this result to understand how the force depends on the
three length scales in the problem: the reduced de Broglie wavelength
$\bar{\la}_\text{dB}$, the skin depth $L_\text{skin} = (f/n_\psi)^{1/2}$,
and the size of the target $R$.
We first consider the limit
\[
L_\text{skin} \ll \bar{\la}_\text{dB} \ll R.
\]
The first inequality is equivalent to $V_\text{eff} \gg k^2/2m_\phi$,
the strong potential regime.
Because $\bar{\la}_\text{dB} \ll R$, we are in the classical
geometric limit where the we can compute the force assuming
that the dark matter is made of classical particles that
scatter elastically from the target.
This is the case where the naive picture used to derive
the estimate \Eq{Fmax} for the force is quantitatively correct.
For example, for a spherical target of radius $R$, we have
\[
F_\text{target} =  \rho_\phi v_\phi^2 \pi R^2.
\]

Next we consider the regime
\[
L_\text{skin} \ll R \ll \bar{\la}_\text{dB}.
\]
We have $kR \ll 1$, so we have $s$-wave scattering,
and we are still in the strong potential regime.
For a spherical target of constant density, 
we have the textbook quantum `hard sphere scattering' problem.
The relevant phase shifts are given by
$\de_0 = -kR$, $\de_1 = O\big( (kR)^3 \big)$, 
and the force is given by
\[
F_\text{target} = 4\rho_\phi v_\phi^2 \pi R^2 .
\]
Once again, this agrees with the estimate \Eq{Fmax}.
Note that the force is 4 times larger than the classical
force in this limit; this is the famous
factor of 4 enhancement of the $s$-wave cross section
compared to classical geometric scattering.

Next we consider the regime
\[
R \ll L_\text{skin} \ll \bar{\la}_\text{dB}.
\]
Because we have $R \ll L_\text{skin}$, the target does not 
strongly affect the incoming wave, and we can use the Born approximation.
In this case, the scattering cross section and the force
are  proportional to $V_\text{eff}^2 \propto 1/f^2$.
Therefore, the scattering is highly suppressed compared to the
strong potential regime, where the cross section is independent
of $f$.
For a sphere of radius $R$
with $kR \ll 1$, the Born scattering amplitude
is given by
\[
f(\th) = -\frac{m_\phi}{2\pi} \myint d^3 r\ggap e^{i \vec{q} \cdot \vec{r}}
\ggap V(\vec{r}) = -\frac{n_\psi R^3}{3f},
\]
where $q = 2k\sin(\th/2)$ is the momentum transfer.
The force is then given by
\[
F_\text{target} = n_\phi v_\phi \int_{-1}^1 d\!\gap\cos\th\,
\frac{1}{2\pi} |f(\th)|^2 \, m_\phi v_\phi (1 - \cos\th)
= 4\pi R^2 \rho_\phi v_\phi^2
\left( \frac{n_\psi R^2}{3f} \right)^2.
\]
Note that the ratio to the maximal force is of order
$(R/L_\text{skin})^4 \ll 1$.

In the weak potential regime $V_\text{eff} \ll k^2/2m_\phi$ 
(or $L_\text{skin} \gg \bar{\la}_\text{dB}$)
we can again use the Born approximation, and the force
is again suppressed by $1/f^2$.
We summarize these results for a general target of size
$R$ in Table~\ref{tab:limit}.
To illustrate this, in Fig.~\ref{fig:Rz} 
we plot the ratio $\mathcal{R}_z$
between the force on a solid aluminum sphere and the
classical particle limit where each dark matter particle
collides elastically with the sphere.

\begin{table}[t]
\centering
\begin{tabular}{|c|c|c|c|c|c|}
\hline
Parametric limit
& $F/F_\text{classical}$     \\
     \hline\hline
   $L_\text{skin}  \ll  \bar{\la}_\text{dB} \ll R$  & 1 \\
  \hline
   $L_\text{skin} \ll R \ll  \bar{\la}_\text{dB}$ &  $\sim 1$ \\ 
      \hline
   $R  \ll L_\text{skin}   \ll  \bar{\la}_\text{dB}  $  & 
   $\sim (R/L_\text{skin})^4$ 
   \\ 
   \hline
$L_\text{skin} \gg \bar{\la}_\text{dB}$ 
&   $\sim ( \text{min}\{ R,  \bar\la_\text{dB}\} / L_\text{skin})^4$ 
\\    \hline
\end{tabular}
\begin{minipage}[t]{5.5in}
\medskip
\caption{Various parametric limits 
of the drag force on a target of linear size $R$.
Here $F/F_\text{classical}$ is the ratio of the dark matter force 
to the force for classical elastic scattering from the target.
The first three entries are for the strong scattering limit
($L_\text{skin} \ll \bar{\la}_\text{dB}$) while the last is
for the weak scattering limit.
All estimates are in the coherent scattering limit where
the de Broglie wavelength of the dark
matter is large enough that we can approximate the target
as a continuous matter distribution.}
\label{tab:limit}
\end{minipage}
\end{table}

\begin{figure}[tb]
\centering
\includegraphics[width=0.6\textwidth]{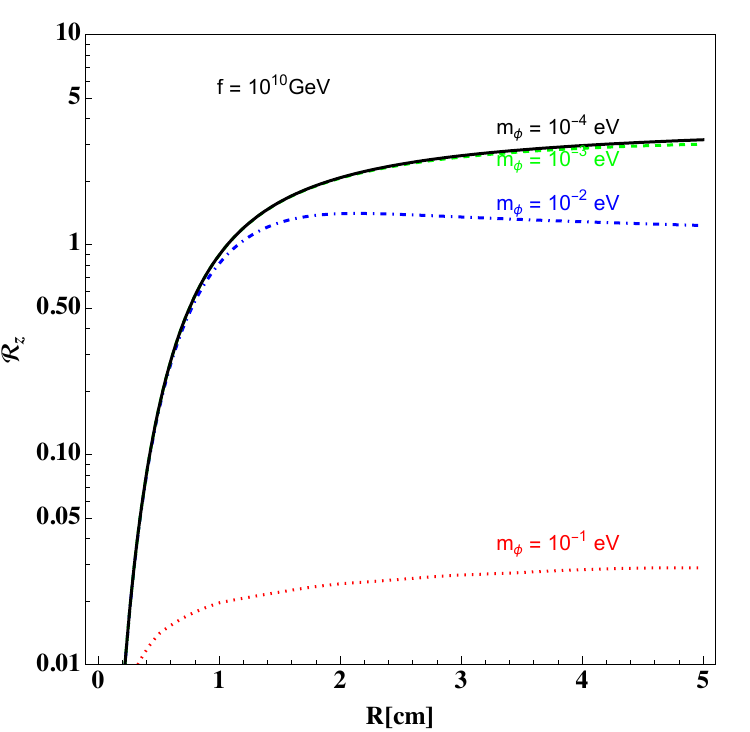}
\\
\begin{minipage}[t]{5.5in}
\caption{Ratio  $\mathcal{R}_z$ between the force 
on a solid sphere of aluminum of radius $R$
and the classical particle limit.
We fix 
$f =10^{10}$ GeV, corresponding to a 
skin depth $L_{\rm skin} = 0.56$~cm, and we have plotted the ratio 
for 
$m_\phi = (10^{-1},10^{-2},10^{-3},10^{-4})$~eV, which corresponds to the reduced de Broglie wavelength $\bar \lambda_{\rm dB} = (0.2, 2.0, 20, 200)$~cm.}
\label{fig:Rz}
\end{minipage}
\end{figure}

\subsection{Shielding Effects}
\scl{shielding}
As we have already stated, ordinary matter can shield test masses
from the dark matter wind.
We now give some quantitative estimates of this effect. 
We focus on 
the strong potential regime
$L_\text{skin} \ll \bar{\la}_\text{dB}$
where shielding can be significant.

We first consider a simple geometry where a plane wave of dark
matter in vacuum with momentum $\vec{k} = k_z \hat{\vec z} + k_y \hat{\vec{y}}$
is incident on an infinite solid plane of thickness $\De R$
occupying the region $0 < z < \De R$.
The steady state solution for the wavefunction is
\[
\psi(\vec{r}) = \begin{cases}
e^{-i(\om t - k_y y)} \big[ e^{i k_z z} 
+ R \gap e^{-i k_z z} \big] & z < 0,
\\
e^{-i(\om t - k_y y)} \big[ A \cosh(\ka_z z) + B\sinh(\ka_z z) \big]
& 0 < z < \De R,
\\
e^{-i(\om t - k_y y)} \times T e^{i k_z z} & z > \De R,
\end{cases}
\]
where 
\[
\ka_z^2 = \frac{n}{f} - k_z^2 > 0.
\]
Matching the wavefunction and its first derivative at the boundaries
determines the unknown coefficients $R$, $A$, $B$, and $T$, and 
gives the for the transmission probability
\[
|T|^2 = \frac{4 k_z^2 \ka_z^2 }{(k_z^2 - \ka_z^2)^2 \sinh^2(\ka_z L) + 4 k_z^2 \ka_z^2 
\cosh^2(\ka_z L)}.
\]
For a plane wave in the $z$ direction we have $\ka = 1/L_\text{skin}$,
and we obtain
\[
\eql{Rinfinitewall}
|T|^2 \simeq \begin{cases}
\displaystyle \frac{16 L_\text{skin}^2}{\bar{\la}_\text{dB}^2}
e^{-2\De R/L_\text{skin}} & \De R \gsim L_\text{skin}, 
\\[12pt]
\displaystyle 1 - \frac{\bar{\la}_\text{dB}^2 \De R^2}{4 L_\text{skin}^4}
& \De R \ll L_\text{skin}.
\end{cases}
\]
%
As expected, the transmission coefficient is exponentially suppressed for
$\De R \gsim L_\text{skin}$, and is
is close to 1 if $\De R \ll L_\text{skin}$. 
This result can be used to approximate the shielding effects if the shield 
can be approximated as a plane on the scale of
the reduced de Broglie wavelength $\bar{\la}_{\text{dB}}$.

We comment briefly on how these results are affected by averaging over
the dark matter distribution in our galaxy.
We use the standard halo model, where the dark matter has a Maxwellian
distribution cut off by the escape velocity:
\[
f(\vec{u}) \propto e^{-2u^2/3\si_u^2} \th(u_\text{esc} - u),
\]
where $\vec{u}$ is the dark matter velocity in the galactic rest frame.
We use $\si_u = 290$~km/s, $u_\text{esc} = 550$~km/s.
We are interested in the dark matter force on an object at rest in the solar system
where the average dark matter wind velocity is 
$v_\text{wind} \simeq 220$~km/s.
This is proportional to
\[
\avg{v_z^2} = v_\text{wind}^2 + \avg{u^2}
\simeq (270~\text{km}/\text{s})^2.
\]
The averaging therefore increases the force by a factor of
roughly $1.5$.
The force exerted by the dark matter on the far side of the
barrier is reduced by the ratio
\[
\eql{transmissionratio}
\scr{T}_z = \frac{\avg{v_z^2 |T|^2}}{\avg{v_z^2}}.
\]
This is illustrated in Fig.~\ref{fig:transmi} along with the approximation where
$|T|^2$ is computed for a pure plane wave in the $z$ direction
with velocity $v_\text{wind}$.
We see that the averaging reduces the effect of shielding.
These effects are relevant for detailed predictions for experiments,
but will not be included in the present exploratory work.

\begin{figure}[!tb]
\centering
\includegraphics[width=0.6\textwidth]{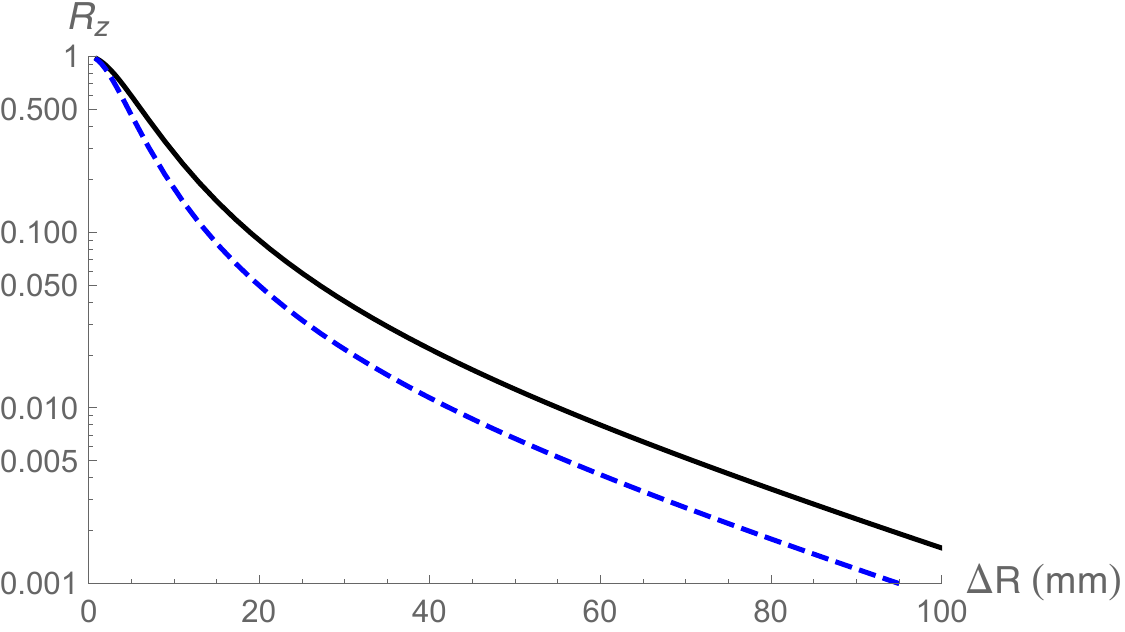}
\\
\begin{minipage}[t]{5.5in}
\caption{The ratio $\scr{R}_z$ of the dark matter force on either side
of a 1-dimensional aluminum barrier of thickness $\De R$
that is normal to the average direction of the dark matter wind
(see \Eq{transmissionratio}).
We take $m_\phi = 10^{-4}$~eV and $f = 10^{12}$~GeV,
which imples $L_\text{skin} = 5.6$~cm.
The black solid line is the result of averaging over the dark matter
velocity distribution in the standard halo model,
while the dashed blue line is the result for a pure plane
wave with velocity 220~km/s.
}.
\label{fig:transmi}
\end{minipage}
\end{figure}

\begin{figure}[!tb]
\centering
\includegraphics[width=0.6\textwidth]{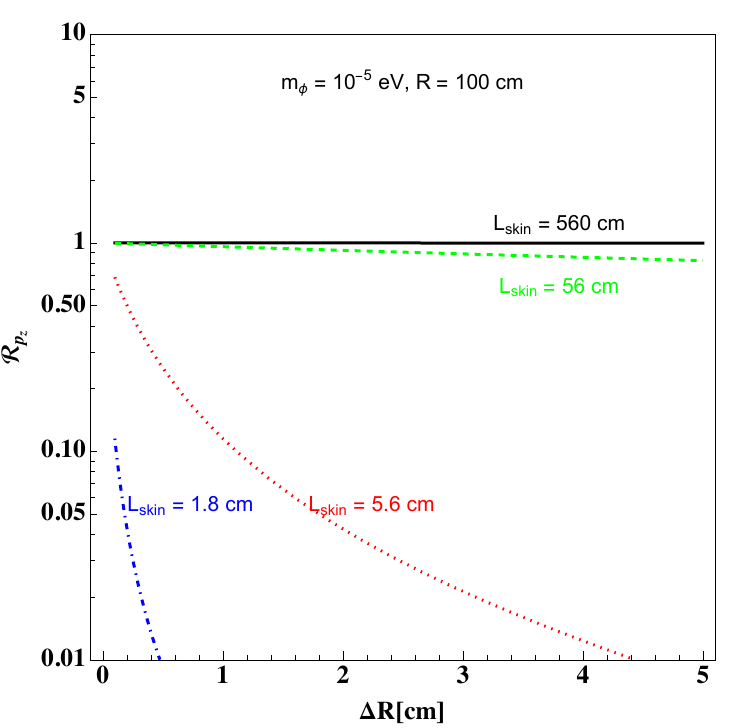}
\\
\begin{minipage}[t]{5.5in}
\caption{Ratio  $\mathcal{R}_{p_z}$ between the $z$-component of the momentum density inside a hollow sphere and its classical limit as a function of the thickness of an aluminum shell with outer radius R = 100 cm. We have fixed $m_\phi =10^{-5}$ eV (i.e. $\bar{\lambda}_\text{dB} = 2 \times 10^3$ cm) and plotted the ratio with $f = (10^{11},10^{12},10^{14},10^{16})$ GeV.}
\label{fig:Rpz}
\end{minipage}
\end{figure}

If $\bar{\la}_\text{dB} \gg R$, 
the shield cannot be approximated by a plane.
This limit 
is relevant only for the very lightest dark matter masses in
the phenomenological regime of interest.
To investigate this regime, we consider a shield consisting of
a spherical shell of of radius $R$ and thickness $\De R \ll R$,
surrounding a test mass at the center.
%
%
To quantify the importance of the shielding, we compute the
momentum density of the dark matter at the center of the sphere,
$\avg{T^{0z}}_{r = 0}$.
This gives the strength of the dark matter `wind' seen by a target
at the center of the sphere.
The solution is given in Appendix A (see \Eq{T0zhollow}). 
In the limit of large $\bar{\la}_\text{dB}$,
the solution is dominated by the
lowest partial wave. 
The ratio $\scr{R}$ of the momentum density at the center 
to the unshielded value is given by
\[
\label{eq:shellmd}
\scr{R}_z = \frac{\avg{T^{0z}}_{r=0}}{\avg{T^{0z}}_{r\to\infty}}
\simeq \Re(D_0^* D\sub{1}),
\]
where the coefficients $D_\ell$ are given in \Eq{ClDlhollow}. 
In the limit
\[
L_\text{skin}  \ll \bar{\la}_\text{dB}, \qquad R \ll \bar{\la}_\text{dB},
\]
we find 
\[
\eql{Rhollowsphere}
\scr{R}_z 
\simeq
\begin{cases}
\displaystyle
\frac{12 L_\text{skin}^2}{R^2} 
e^{-2\De R/L_\text{skin}} 
& \De R \gsim L_\text{skin},
\\[10pt]
\displaystyle 1 - \frac{4}{3} \frac{R \De R}{L_\text{skin}^2}.
& \De R \ll L_\text{skin},
\end{cases}
\]
As we expect, the result is exponentially suppressed for
$\De R \gsim L_\text{skin}$, and near 1 for
$\De R \ll L_\text{skin}$.
However, the coefficients differ from the infinite wall
limit as in \Eq{Rinfinitewall}.
Note that for 
$\De R \ll L_\text{skin}$, the deviation from 1 is of
order $1/L_\text{skin}^2 \sim 1/f$ rather than $1/f^2$ because
it results from the interference of a weakly scattered wave
and the unscattered wave.
For the infinite wall, the the deviation from 1 is
of order $1/L_\text{skin}^4 \sim 1/f^2$ because
unitarity dictates that $|T|^2 + |R|^2 = 1$.

In Fig.~2 we show how the ratio $\scr{R}_z$ depends on the thickness
of the shielding sphere.
The parameters are chosen to approximately match the satellite 
experiment proposed in Ref.~\cite{Buscaino:2015fya}, 
which will be discussed
in  \S\ref{sec:experiments}  below.

\section{Existing Constraints}
\label{sec:constraints}

In this section, we summarize constraints on the dark matter
model from existing observations.
Much of this section is a summary of previous work,
but we also consider additional effects related to the dark Meissner
effect that have not been previously considered in the literature;
we find that these effects do not affect the existing
constraints.


For definiteness, we consider the constraints a benchmark models 
with effective couplings to nucleons and electrons given by
\[
\eql{fNfe}
\scr{L}_\text{eff} = -
\left[ \frac{1}{f_n} \bar{n} n + \frac{1}{f_p} \bar{p} p
+ \frac{1}{f_e} \bar{e} e \right] \sfrac 12 \phi^2.
\]
For most purposes, we can assume that there is an approximately
equal coupling to both protons and neutrons 
with coefficient $f_N^{-1} = \frac 12 (f_n^{-1} + f_p^{-1})$.

\subsection{Supernova Cooling}
The first constraint we consider is the cooling of stars due to $\phi$
emission.
Our model contains an irrelevant interaction,  so the strongest
astrophysical cooling constraint comes from supernova SN1987A, since this
has the highest relevant temperature scale ($T_\text{SN} \sim 30$~MeV).
The constraint can be approximated using the 
`Raffelt criterion'~\cite{Raffelt:1990yz}, which states that the instantaneous 
luminosity for new light particles with effective masses smaller than 
$T_\text{SN}$ cannot exceed the neutrino luminosity observed by the SN1987A. 

For the nucleon coupling, this constraint was estimated in 
\cite{Olive:2007aj}, and gives
\[
f_N \gsim 1.1 \times 10^8~\text{GeV}.
\]
The bound for electron couplings does not appear in the literature,
so we derived it using the approximations described in \cite{Olive:2007aj}.
We constrain the production rate by
\beq
\Gamma(e^+ e^- \rightarrow \phi \phi) \simeq
\frac{7 \zeta(3) T_\text{SN}^7}{320\pi f_e^2} \lesssim10^{-14} \text{ MeV}^{5}
\eeq
which leads to
\beq
\label{eq:febound}
f_e \gsim 1.4\times 10^8 \, \rm GeV.
\eeq
Because of the high density of the supernova core, the mass of the
$\phi$ particles inside the core is much larger than the mass in vacuum.
However, this does not affect the bounds in this model
because the mass is still small compared to the temperature $T_\text{SN}$.%
\footnote{Models with additional contributions to the mass of
light particles inside stars that evade cooling constraints
were considered in \cite{DeRocco:2020xdt}.}
For example, for couplings to nucleons, we have
\[
\sqrt{\De m_\phi^2} = \sqrt{\frac{n_N}{f_N}} \sim 4~\text{keV} 
\left( \frac{f_N}{10^8~\text{GeV}} \right)^{-1/2},
\]
where we assumed $n_N \sim 2 \times 10^{38}/$cm$^3$.
The effects for the electron coupling are even weaker, since
$n_e/n_N \sim 10^{-2}$.

\subsection{Big Bang Nucleosynthesis}
Next we consider the constraints from big bang nucleosynthesis.
During nucleosynthesis $\avg{\phi^2}$ was larger than it is today, and
this  affects the proton-neutron mass difference
and the electron mass via the couplings \Eq{fNfe}.
The relic abundances of nuclei are very sensitive to these quantities,
so this puts a bound on the parameters of the model.
The nucleon and electron mass modification is given by
\[
\label{eq:deltamphi}
\De m_\psi = \frac{\avg{\phi^2}}{2 f_\psi} = \frac{\rho_\phi}{m_\phi^2 f_\psi},
\qquad
\psi = n,\,  p,\, e.
\]
where the dark matter density $\rho_\phi$ is fixed by the cosmological 
evolution.
Therefore, nucleosynthesis 
primarily puts a bound on the parameter combination $1/m_\phi^2 f_\psi$.

These bounds were first obtained 
in \cite{Stadnik:2015kia}, and 
have been refined in 
\cite{Sibiryakov:2020eir,Bouley:2022eer}.
For electron couplings, the constraint in the parameter region of
interest to us can be summarized as
\[
\frac{1}{m_\phi^2 f_e} \lsim 9~\text{GeV}^{-1}\, \text{eV}^{-2}.
\]
For the nucleon couplings, the bounds are model-dependent:
they depend on the form of the couplings
of $\phi$ to quarks and gluons above the QCD confinement scale.
The reason is that the nucleosynthesis bounds are primarily
sensitive to the neutron-proton mass difference,
while the matter effects we are considering in this paper
are primarily sensitive to the sum of the proton and
nucleon couplings.
To illustrate the range of possibilities, we consider
two benchmark models, one where the $\phi$ field couples
only to the down quark, and the second where it couples
only to gluons:
\[
\scr{L}_\text{int} = -\frac{1}{2 \La_d} \phi^2 \bar{d} d
\qquad\text{or}\qquad
-\frac{1}{2 \La_G^2} \phi^2 G^{\mu\nu}_a G_{\mu\nu a}.
\]
Each of these models can be approximately realized by
specific UV completions of the model, as discussed in
Appendix B.
The nucleosynthesis bounds are weaker for the second
model because the gluon coupling contributes
to the neutron-proton mass difference only through small
isospin-breaking effects. In both models, we have $f_p \simeq f_n \simeq f_N$,
and the respective bounds are
\[
\frac{1}{m_\phi^2 f_N} 
\lsim 2 \times 10^{-4}~\text{GeV}^{-1}\, \text{eV}^{-2}
\qquad\text{or}\qquad
\frac{1}{m_\phi^2 f_N} 
\lsim 2 \times 10^{-2}~\text{GeV}^{-1}\, \text{eV}^{-2}.
\]
These constraints are illustrated in Fig.~3 along with the constraints
from supernova cooling.

\begin{figure}[tb]
\centering
\includegraphics[width=0.45\textwidth]{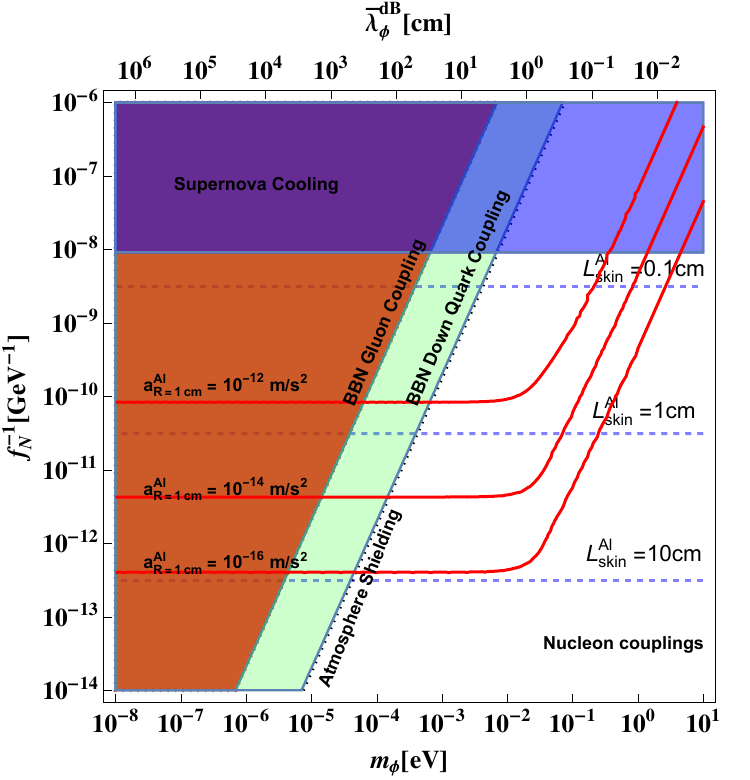}
\includegraphics[width=0.45\textwidth]{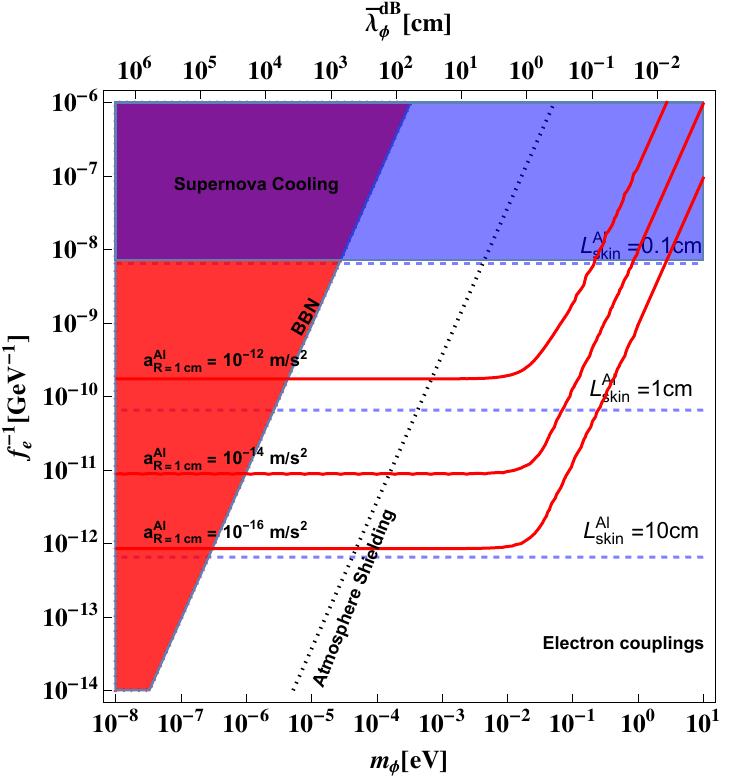}
\begin{minipage}[t]{5.5in}
\caption{Summary of constraints
for couplings of $\phi$ to nucleons (left panel) and electrons (right panel). 
The blue regions are excluded by supernova cooling,
while the red and green regions are excluded by nucleosynthesis.
In the left figure, the red region corresponds to a coupling to gluons,
while the green region corresponds to the coupling to the down quark.
Above the black dotted line, the dark matter wind is shielded by the
earth's atmosphere.
The red solid lines give the acceleration of a solid aluminum target
with radius 1~cm in the absence of shielding.
For comparison, the horizontal blue dashed lines give the
skin depth $L_\text{skin}$ for aluminum.}
\label{fig:constraints}
\end{minipage}
\end{figure}

\subsection{The Dark Drag Force}

Because of the dark matter Meissner effect,  moving ordinary baryonic matter object
will experience a force in the dark matter rest frame
that tends to make it come to rest relative to the dark matter.
For obvious reasons, we call this the dark drag force.
In this subsection, we consider the effect of this force on
ordinary matter objects in a galaxy such as the Milky way.

Our galaxy consists of a dark matter halo, with ordinary matter orbiting
inside the halo with speed $\sim v_\phi$. 
The dark drag force
tends to make baryonic matter come to rest relative to the halo,
possibly modifying galactic dynamics in an observable way.
%
For sufficiently large objects, the collective effects become
important for the drag force.
To estimate this maximum size of this effect, we assume a maximal
acceleration given by \Eq{amax}.
For a given density $\rho$ of the object, the force is proportional
to the area, while the mass is proportional to the volume, 
so the acceleration is proportional to $1/R$,
where $R$ is the size of the object.
The dark drag force is then
large enough to slow the object over the lifetime of the galaxy for
\[
R \lsim \frac{\rho_\phi v_\phi T_\text{gal}}{\rho}
\sim \text{cm} \left( \frac{\rho}{10~\text{g}/\text{cm}^3} \right)^{-1},
\]
where we have normalized $\rho$
to the density of ordinary matter~\footnote{We assume that there are no other non-gravitational forces acting on this object.}. 
(Note that the average density of the sun and Jupiter are both
$\sim 1$~g$/$cm$^3$, which is not so different.)
Such small objects do not play an important role in the dynamics
of the galaxy.
It is intriguing that small chunks of ice (for example)
cannot freely orbit our galaxy, but we know of no observational
constraint arising from this effect.

\section{Detecting the Dark Matter Wind}
\label{sec:experiments}

\begin{figure}[tb]
\centering
\includegraphics[width=0.45\textwidth]{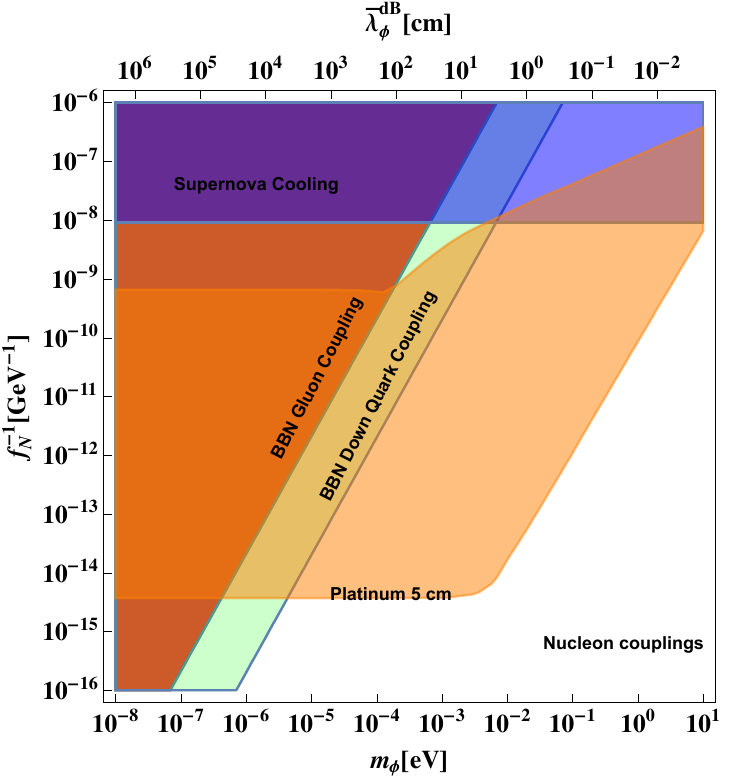}
\includegraphics[width=0.45\textwidth]{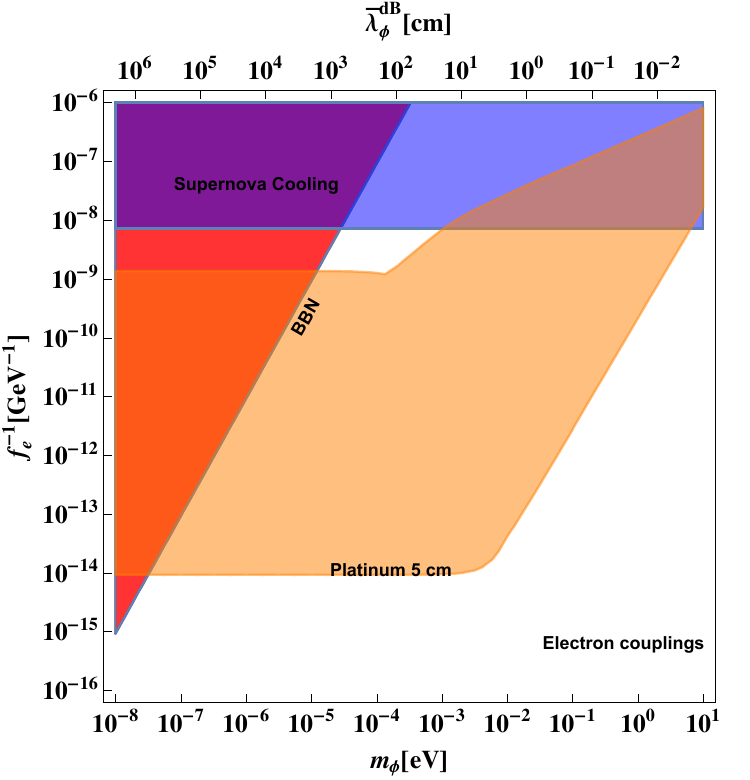}
\begin{minipage}[t]{5.5in}
\caption{Sensitivity on our parameter space 
from the proposed space mission test of modification of gravity
at distances $\sim 10$~AU~\cite{Buscaino:2015fya},
for the nucleon coupling (left panel) 
and the electron coupling (right panel).  
We have assumed a ranging accuracy of 10~cm, over the 7 year lifetime of the
mission, corresponding to an accuracy on the measurement of the acceleration
of $\de a \sim 4 \times 10^{-18}$~m$/$s$^2$.
}
\label{fig:prospect}
\end{minipage}
\end{figure}

In this section we discuss experimental sensitivity to the dark 
matter force.
We first discuss the magnitude of the force neglecting shielding
effects.
We then consider a number of existing force experiments and 
explain why they are not sensitive due to shielding effects.
We then show that the proposed space mission described in
\Ref{Buscaino:2015fya} is sensitive.
Finally, we give a general discussion of some aspects of the
signal that may be relevant for new experiments.

\subsection{Force without Shielding}
To illustrate the magnitude of the force neglecting shielding effects,
we plot the
acceleration of a 10~g spherical aluminum target with radius of 1 cm in Fig.~\ref{fig:constraints},
along with the constraints on the model discussed in \S\ref{sec:constraints}.
For comparison, the sensitivity of the E\"ot-Wash torsion balance
experiment for a similar target ($m \simeq 5$~g)
is $\de a \simeq 9 \times 10^{-15}$~m$/$s$^2$ \cite{Schlamminger:2007ht,Wagner:2012ui}.
The Microscope satellite experiment, which currently puts the strongest 
limits on violations of the equivalence principle, has a sensitivity
of $\de a \simeq 5 \times 10^{-14}$~m$/$s$^2$, but for a larger 
target ($m \sim 300$~g) \cite{MICROSCOPE:2022doy}.
The upgraded version of this experiment (Galileo) is expected to increase
the sensitivity by another 2 orders of magnitude~\cite{Nobili:2012uj} .
We see that  without any shielding, the accelerations that can be detected in existing or
planned weak force experiments are
nominally sensitive to the dark matter wind
a wide range of parameter space of the model. 

\subsection{Existing Experiments}
We consider force 
experiments that search for violations of the equivalence
principle~\cite{Wagner:2012ui,MICROSCOPE:2022doy,Nobili:2012uj},
since these typically
involve very precise measurements of forces on large
(cm scale) targets.
The gravitational acceleration is independent of the target, while
the acceleration produced by the dark matter wind is proportional to
the area of the target and inversely proportional to its mass.
In this sense, the dark matter wind induces a violation of the
equivalence principle.

However, to 
estimate the signal, we must take into account shielding effects.
%
For example, terrestrial experiments will be sensitive only if the dark matter
can penetrate the earth's atmosphere.
The atmosphere will block dark matter wind
if $\De m_\phi^2 \sim n_\text{atm}/f \gsim m_\phi^2 v_\phi^2$, or
\[
\frac{f^{-1}}{m_\phi^2} \gsim 10^{-4}
~\text{GeV}^{-1} \, \text{eV}^{-2}.
\]
This is the region above the dashed black line in Fig.~\ref{fig:constraints}.  The effects of shielding in an experiment cannot be determined 
quantitatively without detailed
modeling the experiment and its environment.
Rough estimates for existing experiments
indicate that the shielding
is too large for them to be sensitive to the dark matter wind.
For example, torsion balance experiments such as the E\"ot-Wash experiment \cite{Schlamminger:2007ht,Wagner:2012ui} are performed in a 
vibration-isolated laboratory surrounded by meters of dense matter.
The probe masses in the Microscope experiment \cite{MICROSCOPE:2022doy}
are surrounded by instrumentation
with thickness $\sim 5$~cm, in addition to the  shielding due to the surrounding satellite.
Simple estimates of the shielding based on simplified geometries
such as the ones discussed in \S\ref{sec:shielding} indicate that these experiments
are not sensitive to the dark matter wind.
The lesson is that future experiments will need to be carefully
designed to ensure that shielding effects are small.

\subsection{Spacecraft Ranging}
One proposed experiment that is 
sensitive to the dark matter wind is the space mission 
proposed in Ref.~\cite{Buscaino:2015fya}.
This experiment was designed to test the inverse square law of the 
gravitational force on a distance scale of 1--100~AU. 
The spacecraft is sent out of the orbital plane of the solar system,
where the gravitational force is dominated by the sun.
The force measurement is made by a 
`drag free' spacecraft that steers
around a proof mass floating in its center.
A second spacecraft $\sim 10$~km away sends ranging information
back to earth to measure the distance from the proof mass to
an earth station with an accuracy in the range 10--100~cm.
The expected sensitivity to a deviation in the radial acceleration
assuming an optimistic ranging accuracy of $10$~cm
is $\de a \sim 4 \times 10^{-18}$~m$/$s$^2$.
The sensitivity benefits from the fact that the deviation from
the expected geodesic orbit builds up over the 7 year run time
of the experiment.
The drag-free spacecraft has 
a simple spherical geometry to control various background effects.
Using the parameters in Ref.~\cite{Buscaino:2015fya} we approximate the
satellite as a spherical aluminum shell of radius 1~m and thickness 1~mm.
The proof mass in the design is a solid platinum 
sphere of radius 5~cm, with a mass of $\sim 10$~kg.
In principle the directionality of the dark matter force could be
used to distinguish it from a violation of the inverse-square
law for the gravitational force, but this experiment is
sensitive only to the distance to the earth, and hence
the component of the acceleration along this direction.
In Fig.~\ref{fig:prospect} we give the sensitivity to the dark matter wind for the proposed
experimental parameters. 
The force on the proof mass is estimated as follows.
For $m_\phi \lsim 10^{-4}$~eV, the de Broglie wavelength is
large compared to the size of the spacecraft, and we use the 
result of \Eq{shellmd} for the momentum density inside a
spherical shell.
For $m_\phi \gsim 10^{-4}$~eV we approximate the shielding
effect by approximating the shielding by a planar wall.
In the intermediate region, we simply interpolate.%
\footnote{A more realistic approach is to calculate the force 
directly on a solid sphere inside a spherical shell.
Some results on this are presented Appendix~\ref{app:sphereinshell}.
However, these results are challenging to implement numerically, and we
did not use them to generate the plots.}
%
We see that the experiment is sensitive to a large part of the
parameter space due to a combination of minimal shielding and
high acceleration sensitivity.

\subsection{General Signal Characteristics}
There are a number of characteristics of the signal that could
be used to design other experiments sensitive to the dark matter wind.
First, we note that the dark matter wind is known to be coming from
the direction of
the constellation Cygnus, which is
visible in the northern hemisphere.
Terrestrial experiments therefore have a daily modulation in
the direction of the force, including the disappearance of the
force when Cygnus is below the horizon.
The reflection of the dark matter from the surface of the
earth means that the net direction of the wind 
is horizontal in the approximation
where the only matter near the experiment is the (nearly) flat
surface of the earth.
However, the dark matter wind will be sensitive to other
dense objects nearby, and modeling of the interaction of the dark matter
with the experimental environment will be required to determine
if there is a signal whose characteristics can be understood.
For satellite-based experiments, there is a modulation once per
orbit due to the orbit of the satellite.
The earth will have a dark matter shadow where the force is suppressed,
and reflection from the earth's surface must be taken into account.

In addition, both terrestrial and satellite experiments will
experience an annual modulation of the signal due to the
earth's motion around the sun.
The earth orbits the sun with a speed of approximately 30~km$/$s,
approximately $10\%$ of the average speed of the dark matter wind.
The dark matter force is proportional to the square of the wind
velocity (see \Eq{Fmax}), so this will give rise to a significant
modulation in the magnitude of the force.
The phase and magnitude of this modulation is known, 
and provides an additional handle on the signal.


\section{Conclusions}
\label{sec:conclusion}
In this paper, we have explored a model of dark matter where the interaction
between dark matter and ordinary matter can be maximally strong,
in the sense that dark matter scatters elastically from sufficiently
dense and large matter targets.
The dark matter in this model consists of a scalar $\phi$
with mass $m_\phi \lsim \text{eV}$ and
an effective coupling to nucleons and/or electrons given by \Eq{theint}.
This coupling increases the mass of the $\phi$ particle inside
ordinary matter, which suppresses the propagation of dark matter
inside the target and can lead to elastic scattering.
This is a collective effect due to the coherent scattering of the
dark matter from many nucleons/electrons.
It is similar to the Meissner effect that gives photons
a mass inside a superconductor, so we call it the `dark Meissner effect.'

Because of the rotation of the Milky Way inside the dark matter halo,
the dark matter around the earth is moving with an average velocity
$v_\phi \sim 300$~km$/$s from the direction of Cygnus.
The maximal force from this dark matter `wind' that arises for
elastic scattering is very small (see \Eq{amax}), but is large enough
to be detected in sensitive force measurements.
However, the strong interaction between dark matter and ordinary
matter also means that the dark matter is shielded from existing
experiments.

We have shown how this force can be computed quantitatively,
using both the classical and quantum pictures for the dark matter.
We present a number of explicit calculations to illustrate how the
force depends on the dark matter de Broglie wavelength,
target size and density, amount of shielding, {\it etc\/}.
These calculations confirm that the dark matter force can
indeed be maximal in a wide range of parameters.
Based on these estimates, we believe that existing fifth force
experiments (both terrestrial and satellite based)
are not sensitive to the dark matter force because of shielding
effects. We also show that this 
model is consistent with astrophysical  and cosmological
constraints in a region where the induced acceleration on the test object with radius of  $~$cm is within the sensitivity of current technology  as long as the shielding effect can be reduced (see Fig.~\ref{fig:constraints}).

This leads us to consider a novel experimental signal of dark matter:
multiple elastic scatterings of dark matter from ordinary matter that
are accumulated during a long period of time, leading to a collective
force that can be detected using sensitive force experiments.
This is a unique signal with many characteristics that can be
used to distinguish it from backgrounds:
the force has a known direction, annual modulation due to the
earth's orbit, and time dependence due to the earth's shadow.
Perhaps most uniquely, the force is proportional to the area
of the target, and dark matter can be shielded and/or controlled
by ordinary matter. 
Detecting such a signal would not only give direct evidence of 
dark matter, but also information about its local velocity
distribution.

To detect this signal, one needs a sufficiently sensitive force
probe that is not too shielded from the dark matter wind.
The size of the force to be measured is larger than forces
already probed in these experiments although special design should be explored to reduce the shielding effect, for example thinner shielding made with low density materials. As an example, we have presented the prospective sensitivity on our parameter space (see Fig.~\ref{fig:prospect}) from proposed space mission test of long-distance modification of the gravity inverse-square law by Ref.~\cite{Buscaino:2015fya}, which consists of an aluminum shell of thickness 1 mm with radius 1m.
We hope that our results will stimulate work in this
direction.

\textbf{Note:} As we were completing this work, we became aware of Ref.~\cite{VanTilburg:2024tst}, 
which demonstrates that
the force mediated by the $\phi$ field has a range 
of order the de Broglie wavelength (rather than the Compton wavelength)
due to the presence of a background $\phi$ field.
That work does not give results for the case where the collective effects 
discussed in this paper are important.
Ref.~\cite{VanTilburg:2024tst} also gives a stronger bound on the $\phi^4$ coupling from
the bullet cluster, modifying the tuning estimates in our Appendix B.

\section*{Acknowledgements}
We would like to thank 
Savas Dimopoulos, 
Peter Graham,  
Surjeet Ragendran,
Hari Ramani,
Dam T. Son,
and Ken Van Tilburg
for useful discussions.  
We also thank Ken Van Tilburg for coordinating and sharing with us
his work on other aspects of this model.
The work of HD was partially supported by the UC Davis Physics REU program 
under NSF grant PHY2150515, and by the U.S. Department of Energy 
under grant DE-SC0015655. 
The work of DL was partially supported by the U.S.~Department of Energy 
under grants DE-SC0007914 and DE-SC-0009999, and by
by PITT PACC.
The work of ML was supported by U.S. Department of Energy under grant  
DE-SC-0009999. 
The work of YZ was supported by the U.S. Department of Energy under 
grant DE-SC0009959.
The work of DL and ML
was performed in part at the Aspen Center for Physics, 
which is supported by National Science Foundation grant PHY-2210452.

\appendix{Appendix A: Calculations and Results}
\label{app:detailforce}
In this appendix, we collect additional details about the calculation
of the force due to the dark matter wind.
We also present results of calculations for particular cases.

We are interested in steady-state solutions of the form \Eq{steadyphi}
that describe scattering of an incident wave in the $+z$ direction.
This means that $\psi(\vec{r})$ has the form
\[
\eql{psiscatt}
\psi(\vec{r}) = \phi_0 \Big[
e^{ikz} + \psi_\text{scatt}(\vec{r}) \Big],
\]
where $k = m_\phi v_\phi$.
The classical equations of motion for $\phi$ imply that
$\psi(\vec{r})$ satisfies the
non-relativistic Schr\"odinger equation, \Eqs{NRSE} and \eq{Veff}.
This leads to the equivalence between the force computed in the
classical and quantum pictures, as we will see.

Finding the solution to this problem is a standard problem in
non-relativistic quantum mechanics.
At large distances where the matter density vanishes, the
scattered wave can be written as
\[
\label{eq:psiscattapp}
\psi_\text{scatt}(\vec{r}) = \sum_{\ell \,=\, 0}^\infty
i^\ell (2\ell + 1) P_\ell(\cos\th) A_\ell 
\Big[ j_\ell(kr) + i y_\ell(kr) \Big],
\]
where we  have defined
\[
\label{eq:Aelldefn}
A_\ell = \frac{S_\ell - 1}{2}.
\]
Here $S_\ell$ is the quantum-mechanical $S$-matrix element for the
partial wave $\ell$ (see \Eq{Sell}).
Here $j_\ell$ ($y_\ell$) are the spherical Bessel functions
of the first (second) kind, and $P_\ell$ are the Legendre
polynomials.
Eqs.~(\ref{eq:psiscatt}) and (\ref{eq:psiscattapp}) can be understood from the partial wave expansion of the
full solution:
\[
\psi(\vec{r}) = \phi_0 \sum_\ell
i^\ell (2\ell + 1) C_\ell(r) P_\ell(\cos\th) ,
\]
where the partial waves for $r \to \infty$ are given by
\[
C_\ell(r) = j_\ell(kr)
+ A_\ell  \big[ j_\ell(kr) + i y_\ell(kr) \big]
\to 
\frac{\HD{-}e^{-i(kr - \ell\pi/2)} + S_\ell e^{i(kr - \ell\pi/2)}}{2ikr}.
\]
This corresponds to a phase shift
$S_\ell$ for the outgoing wave.

The solution above describes both the classical $\phi$ field,
and also the wavefunction for a single $\phi$ particle in the quantum
picture.
However, the calculation of the force is different in the two pictures.
In the classical formulation, we compute the force on the target by
computing the momentum transferred to the target by the field.
In the presence of the target, the spatial components of the
energy-momentum tensor of  the $\phi$ field are not conserved:
\[
\d_\mu T^{\mu i}_\phi = 
-\frac{\d_i n(\vec{r})}{2f} \phi^2,
\]
where $T^{\mu\nu}_\phi$ is the energy momentum tensor for the
$\phi$ field
\[
T^{\mu\nu}_\phi = \d^\mu \phi \d^\nu \phi 
- \eta^{\mu\nu}\scr{L}_\phi,
\qquad
\scr{L}_\phi = \frac 12 (\d\phi)^2 - \frac 12 \big(
m_\phi^2 + n(\vec{r})/f \big) \phi^2.
\]
Therefore, the presence of a target transfers momentum to the
field.
The force on the target can therefore be obtained from
the momentum transfer to the field:
\[
\eql{Fstart}
F^i_\text{target} = 
-\frac{d}{dt} P_\phi^i = -\myint dA\, T_\phi^{ij} \hat{n}^j.
\]
The integration surface is taken in a region where $n(\vec{r}) = 0$,
so $T^{\mu\nu}_\phi$ is conserved and
the integral is independent of the choice of surface.
We can therefore evaluate the integral over a sphere with radius
$r \to \infty$.
Because $dA \propto r^2$, the only terms that 
contribute are those where $T^{\mu\nu}_\phi \sim 1/r^2$.

Because we are interested in the force on time scales
$t \gg 1/\om$, we will average over the oscillations in time.
The averaged components of the energy-momentum tensor are given by
\begin{subequations}
\[
\avg{T^{00}_\phi} &= \sfrac 14 \Big[ (\om^2 + m_\phi^2) |\psi|^2
+ |\grad\psi|^2 \Big],
\eql{T00}
\\
\avg{T^{ij}_\phi} &= \avg{T_{\phi\ggap ij}} = \sfrac 14 \Big[
\d_i \psi^* \d_j \psi + \d_j \psi^* \d_i \psi - \sfrac 12 \de_{ij} 
\grad^2 |\psi|^2 \Big],
\eql{Tij}
\]
\end{subequations}
where $i, j$ are Cartesian spatial indices. 
The force on the target is in the $z$ direction, so we can write
\[
\avg{F_\text{target}^z}
&= -\myint dA \ggap \avg{T^{zi}_\phi} \hat{r}^i
\nn
&= -2\pi r^2
\int_{-1}^1 \! d\!\gap\cos\th \,
\sfrac 14 \Big[
\big( \d_z \psi^* \d_r \psi + \hc \big) - \sfrac 12 \grad^2 |\psi|^2 \cos\th
\Big].
\]
Because the integral is over a surface with $r \to \infty$, only
the terms $\sim 1/r^2$ in the integrand contribute.

Keeping only the leading terms in the expansion in $1/r$
in the partial wave expansion
is equivalent to using the approximations
\[
\d_z\psi \simeq \cos\th \d_r \psi,
\qquad
\grad^2 |\psi|^2 \simeq \d_r^2 |\psi|^2,
\]
which gives
\[
\eql{Fzdrpsi}
F_\text{target}^z &= -\frac{\pi r^2}{2} 
\int_{-1}^1 \!d\!\gap\cos\th \,
\Big[
|\d_r \psi|^2 - \sfrac 12 \big(\psi^* \d_r^2 \psi + \hc\big)
\Big] \cos\th + O(1/r).
\]
By using the identities
\[
j'_\ell(\rho) = -j_{\ell+1}(\rho) + \frac{\ell}{\rho} j_\ell(\rho),
\qquad
y'_\ell(\rho) = -y_{\ell+1}(\rho) + \frac{\ell}{\rho} y_\ell(\rho),
\]
and the asymptotic forms of the spherical Bessel functions, we obtain
\footnote{
Taking the derivative inside the partial wave sum as we do here
corresponds to using an IR regulator $\ell < L$ and taking $L \to \infty$
at the end of the calculation.}

\[
(\d_r)^n \psi = -\frac{i\phi_0}{2kr} (ik)^n
\sum_\ell (2\ell + 1) P_\ell(\cos\th)
\Big[ S_\ell e^{ikr} - (-1)^{\ell+n} e^{-ikr} \Big] + O(1/r^2).
\]
The angular average is performed using the identity
\[
\int_{-1}^1 d\!\gap\cos\th \,
P_\ell(\cos\th)
P_{\ell'}(\cos\th) \cos\th
= \begin{cases}
\displaystyle \frac{2\max(\ell, \ell')}{(\ell + \ell')(\ell + \ell' + 2)}
& \ell - \ell' = \pm 1, \\[12pt]
0 & \text{otherwise},
\end{cases}
\]
and we obtain
\[
\eql{Fzappendix}
F^z_\text{target}
= - \frac{\pi}{2} |\phi_0|^2 \sum_\ell (\ell + 1)
\Big[ S_\ell^* S\sub{\ell + 1} + S_{\ell+1}^* S\sub{\ell} - 2 \Big].
\]
With the normalization \Eq{phinorm}, this gives the result \Eq{Fzclassical}
for the force.

We now compare this to the momentum transfer computed in the quantum
mechanics picture.
In this picture, the momentum is transferred to the target by the
scattering of individual $\phi$ particles.
The scattering cross section is determined by the 
$r \to \infty$ behavior of the scattered
wave~\cite{Landau:1991wop}
\[
\psi_\text{scatt}(\vec{r}) \to 
\frac{e^{ikr}}{r} f(\th),
\qquad
f(\th) = \frac{1}{2ik}\sum_\ell (2\ell + 1) P_\ell(\cos\th) (S_\ell - 1).
\]
The differential scattering cross section is given by
\[
\frac{d\si}{d\Om} = \frac{1}{2\pi} \frac{d\si}{d\cos\th} = |f(\th)|^2,
\]
so the momentum transferred to the target is given by
\[
F^z_\text{target} &= n_\phi v_\phi 
\int_{-1}^1 d\!\gap\cos\th\,
\frac{d\si}{d\cos\th} 
\ggap m_\phi v_\phi (1 - \cos\th)
\nn
&= -\frac{\pi \rho_\phi}{m_\phi^2} 
\sum_\ell (\ell + 1)
\Big[ S_\ell^* S\sub{\ell + 1} + S_{\ell+1}^* S\sub{\ell} - 2 \Big],
\]
where we used $k = m_\phi v_\phi$.
This agrees with the classical result \Eq{Fzclassical}.

\subsection{Solid sphere}
\scl{SolidSphere}
We first consider scattering from a sphere of radius $R$ and number
density $n_0$.
This is a standard problem in quantum mechanics.
The wavefunction $\psi(\vec{r})$ outside the sphere is given by the
expansion \Eq{psiscattapp},
while the wavefunction inside is given by
\[
  \psi_{r<R}(\mathbf{r})= \phi_0 \sum_\ell  (2\ell+1) i^\ell B_\ell j_\ell(\tilde{k}r)P_\ell(\cos\th), \label{in}
\]
where
\[
\tilde{k}=\sqrt{k^2- n_0/f\,}.
\]
Note that the wavefunction for $r < R$
does not have a $y_\ell$ term because this is singular at the origin. 
The coefficients $A_\ell$ (which are equivalent to $S_\ell$, see \Eq{Aelldefn})
and $B_\ell$ that define the solution are
obtained by requiring that the wavefunction and its first derivative
are continuous at $r = R$.
The result is
\begin{subequations}
\label{eq:Alsolid}
\[
    A_\ell &=\frac{- k j_\ell(\tilde{k}R) j_\ell^\prime(kR)+\tilde{k} j_\ell^\prime(\tilde{k}R) j_\ell(kR)}{ kj_\ell(\tilde{k}R)h_\ell^{(1)\prime}(kR)-\tilde{k}j_\ell^\prime(\tilde{k}R)h^{(1)}_\ell(kR)},
\\[5pt]
B_\ell &= \frac{kh^\prime_\ell(kR)j_\ell(k R)-kj^\prime_\ell(kR)h_\ell(kR)}{kh^\prime_\ell(kR)j_\ell(\tilde k R)-\tilde k j^\prime_\ell(\tilde k R)h_\ell(kR)}.
\]
\end{subequations}

\subsection{Hollow sphere}
\label{app:hollowsphere}
To model shielding effects, we now
consider a spherical shell with inner radius $R_1$, outer
radius $R_2$, and density $n_0$:
\[
n(r) = n_0 \big[ \th(R_2 - r)-\th(R_1- r)\big].
\]
Here we are interested in computing the 
momentum density $\avg{T^{0z}}$
at the center of the sphere (see Fig.~\ref{fig:Rpz}). We have
\begin{subequations}
\[
  \psi_{r>R_2}(\mathbf{r}) 
  &=\phi_0 \sum_\ell  (2\ell+1) i^\ell \left[j_\ell(kr) + A_\ell h_\ell^{(1)}(kr)\right]P_\ell (\cos\th), \label{outh}
\\
    \psi_{R_1<r<R_2}(\mathbf{r})
    &= \phi_0 \sum_\ell  (2\ell+1) i^\ell \left[B_\ell j_\ell(\tilde{k}r)+C_\ell y_\ell(\tilde{k}r) \right]P_\ell (\cos\th), \label{midh}
\\
  \psi_{r<R_1}(\mathbf{r}) &= \phi_0 \sum_\ell  (2\ell+1) i^\ell  D_\ell j_\ell(kr)P_\ell(\cos\th),
  \label{inh}
\]
\end{subequations}
where $\tilde{k}=\sqrt{k^2-\frac{n}{f}}$.
The wavefunction and its first derivative 
must be continuous at both the inner and outer surfaces,
and we obtain
\begin{subequations}
\label{eq:Alhollow}
\[
A_\ell &=\frac{\tilde{k} j_\ell(kR_2)\left(j_\ell^\prime(\tilde{k}R_2)
+ N_\ell y_\ell^\prime(\tilde{k}R_2)\right)-k j_\ell^\prime(kR_2)\left(j_\ell(\tilde{k}R_2)+N_\ell y_\ell(\tilde{k}R_2)\right)}{k h_\ell^{(1)\prime}(kR_2)\left(j_\ell(\tilde{k}R_2)+N_\ell y_\ell(\tilde{k}R_2)\right)-h_\ell^{(1)}(kR_2)\left(\tilde k j_\ell^\prime(\tilde{k}R_2)+N_\ell  \tilde k y_\ell^\prime(\tilde{k}R_2)\right)},
\\
B_\ell&=\frac{k j_\ell(kR_2) h_\ell^{(1)\prime} (kR_2)-k j_\ell^\prime(kR_2) h_\ell^{(1)}(kR_2)}{k h_\ell^{(1)\prime}(kR_2)\left(j_\ell(\tilde{k}R_2)+N_\ell y_\ell(\tilde{k}R_2)\right)-h_\ell^{(1)}(kR_2)\left(\tilde k j_\ell^\prime(\tilde{k}R_2)+N_\ell  \tilde k y_\ell^\prime(\tilde{k}R_2)\right)},
\\
\label{eq:ClDlhollow}
C_\ell &= B_\ell N_\ell,\quad D_\ell = \frac{B_\ell}{j_\ell(kR_1)} \left[ j_\ell(\tilde{k}R_1)+N_\ell y_\ell(\tilde{k}R_1)\right],
\]
\end{subequations}
where $N_\ell$ is given by
\begin{equation}
N_\ell=\frac{k j_\ell^\prime(kR_1)j_\ell(\tilde{k}R_1)- \tilde k j_\ell^\prime(\tilde{k}R_1)j_\ell(kR_1)}{\tilde k y_\ell^\prime(\tilde{k}R_1)j_\ell(kR_1)- k j_\ell^\prime(kR_1)y_\ell(\tilde{k}R_1)}.
\end{equation}
The momentum density inside the hollow sphere ($r < R_1$) is given by
directly using Eq.~(\ref{eq:Tij}):
\[
\eql{T0zhollow}
\avg{T^{0z}(r, \th)} &= \rho_\phi v_\phi \sum_{\ell \ne \ell^\prime} (2\ell+1)(2 \ell^\prime + 1) \Re\!\big[  i^{\ell^\prime + \ell - 1}  D_\ell^* D\sub{\ell^\prime} \big]  j_\ell( kr ) P_{\ell}(\cos\th)
\nn[-5pt]
&\qquad\qquad\qquad{}
\times \left[j^\prime_{\ell^\prime} (k r)  \cos\th P_{\ell^\prime} (\cos\th) + \frac{j_{\ell^\prime}(kr)  }{k r} \sin^2\th P^\prime_{\ell^\prime} (\cos\th)\right].
\]

\subsection{Sphere in a Shell}
\label{app:sphereinshell}
To determine the force on a sphere that is surrounded by a shield, 
we consider a sphere of radius $R_0$ and density $n_{\text{center}}$, 
and a shell with inner radius $R_1$, outer radius $R_2$, and density $n_{\text{shell}}$. 
In this case, we have 
\begin{equation}
n(r)=n_\text{shell}\left(\Theta(R_2 - r)-\Theta(R_1 - r)\right)+n_\text{center}\Theta(R_0 - r),
\end{equation}
and the system of equations is
\begin{subequations}
\[
  \psi_{r>R_2}(\mathbf{r}) &= \phi_0 \sum_\ell (2\ell+1)i^\ell\left[j_\ell(kr) + A_\ell h_\ell^{(1)}(kr)\right]P_\ell (\cos\theta),\\
  \psi_{R_1<r<R_2}(\mathbf{r})&=\phi_0 \sum_\ell  (2\ell+1) i^\ell \left[B_\ell j_\ell(\tilde{k}_s r)+C_\ell n_\ell(\tilde{k}_s r) \right]P_\ell (\cos\theta),\\
  \psi_{R_0<r<R_1}(\mathbf{r})&= \phi_0 \sum_\ell  (2\ell+1) i^\ell  \left[D_\ell j_\ell(kr)+E_\ell n_\ell(kr) \right]P_\ell (\cos\theta),\\
  \psi_{r<R_0}(\mathbf{r})&=\phi_0 \sum_\ell (2\ell+1)i^\ell F_\ell j_\ell(\tilde{k}_c r)P_\ell(\cos\theta),
\]
\end{subequations}
where $\tilde{k}_s=\sqrt{k^2-\frac{n_\text{shell}}{f}}$ and $\tilde{k}_c=\sqrt{k^2-\frac{n_\text{center}}{f}}$.  The boundary conditions at the surfaces are that the wavefunction and its derivative must be continuous. Thus we must have
\begin{subequations}
\[
A_\ell &=\frac { \tilde k_s j_\ell(kR_2)\left( j_\ell^\prime(\tilde{k}_sR_2)+N_\ell   n_\ell^\prime(\tilde{k}_s R_2)\right)-k j_\ell^\prime(kR_2)\left(j_\ell(\tilde{k}_s R_2)+N_\ell n_\ell(\tilde{k}_s R_2)\right)}{k h_\ell^{(1)\prime}(kR_2)\left(j_\ell(\tilde{k}_s R_2)+N_\ell n_\ell(\tilde{k}_s R_2)\right)- \tilde k_s h_\ell^{(1)}(kR_2)\left( j_\ell^\prime(\tilde{k}_s R_2)+N_\ell  n_\ell^\prime(\tilde{k}_s R_2)\right)}, \\
B_\ell &=\frac{ k j_\ell(kR_2)  h_\ell^{(1)\prime} (kR_2)-k j_\ell^\prime(kR_2) h_\ell^{(1)}(kR_2)}{k h_\ell^{(1)\prime}(kR_2)\left(j_\ell(\tilde{k}_s R_2)+N_\ell n_\ell(\tilde{k}_s R_2)\right)- \tilde k_s  h_\ell^{(1)}(kR_2)\left( j_\ell^\prime(\tilde{k}_s R_2)+N_\ell   n_\ell^\prime(\tilde{k}_s R_2)\right)},\\
C_\ell &= N_\ell B_\ell, \qquad D_\ell = \frac{j_\ell(\tilde{k}_sR_1)+N_\ell n_\ell(\tilde{k}_sR_1)}{j_\ell (k R_1) + M_\ell n_\ell (k R_1)} B_\ell , \\
E_\ell &= D_\ell M_\ell,\qquad  F_\ell = \frac{D_\ell}{j_\ell(\tilde{k}_c R_0)} \left[ j_\ell(kR_0)+M_\ell n_\ell(kR_0)\right],
\]
\end{subequations}
where $M_\ell$ and $N_\ell$ are given by
\begin{equation}
M_\ell=\frac{ k j_\ell(\tilde{k}_c R_0)j_\ell^\prime(kR_0)- \tilde{k}_c j_\ell^\prime(\tilde{k}_cR_0)j_\ell(kR_0)}{\tilde{k}_c n_\ell(kR_0) j_\ell^\prime(\tilde{k}_cR_0)- k n_\ell^\prime(kR_0)j_\ell(\tilde{k}_cR_0)},
\end{equation}
\begin{equation}
N_\ell  = \frac{-\tilde k_s j_\ell^\prime (\tilde k_s R_1) (j_\ell(kR_1) + M_\ell n_\ell (kR_1)) + k j_\ell (\tilde k_s R_1)\left( j_\ell^\prime (kR_1) + M_\ell n_\ell^\prime (kR_1)\right)}{\tilde k_s n_\ell^\prime (\tilde k_s R_1) (j_\ell(kR_1) + M_\ell n_\ell (kR_1)) - k n_\ell (\tilde k_s R_1)\left( j_\ell^\prime (kR_1) + M_\ell  n_\ell^\prime (kR_1)\right)}.
\end{equation}

To calculate the force on the sphere inside the shell, we cannot use the asymptotic solution above (\Eq{Fzappendix}). Instead, we must start at \Eq{Fstart} and perform the calculation without dropping $\mathcal{O}(1/r^2)$ terms. This results in 
\begin{equation}
\begin{aligned}
\eql{forcegeneralsphere}
F_z=\frac{i\pi}{2}\phi_0^2\sum_\ell
(\ell+1)\Big[&r^2\big(f^*_\ell(r)\partial_r^2 f_{\ell+1}(r)-f^*_{\ell+1}(r)\partial_r^2 f_\ell(r)\big)+2r^2\partial_r f_{\ell+1}^*(r)\partial_r f_\ell(r)\\
&{} -2r(\ell+1)\big(\partial_r f_\ell^*(r)f_{\ell+1}(r)+\partial_r f_{\ell+1}^*(r)f_\ell(r)\big)+2f_{\ell+1}^*(r)f_\ell(r)
\\[3pt]
&{} - \text{h.c.}\Big],
\end{aligned}
\end{equation}
where $f_\ell(r)$ is the $r$-dependent component of the wavefunction in the region where the force is applied. In this case, we have 
\begin{equation}
f_\ell(r)=D_\ell j_\ell(kr)+E_\ell y_\ell(kr)
\end{equation}
for the region outside the target sphere
and inside the shell.
%
%
Although the result appears to be $r$-dependent, 
it is independent of $r$ in the region where $f_\ell(r)$ is defined. 
%
\Eq{forcegeneralsphere} holds for any spherical geometry, and we have checked that 
it gives the correct result for the force
on a solid sphere with 
$f_\ell(r)=j_\ell(kr)+A_\ell h_\ell^{(1)}(kr)$ (see \Eq{Alsolid}).
%

\appendix{Appendix B: UV Completions}
\label{app:UVmodel}
In this appendix, we consider two possible UV completions of the effective 
interactions \Eq{theint} and show how they connect with the different
cases for the nucleosynthesis bound discussed in \S\ref{sec:constraints}.
We also briefly discuss the tuning in these models.

As one would expect for a model of a light scalar with non-derivative couplings,
the mass $m_\phi$ is very fine-tuned.
The dark matter relic abundance depends sensitively on $m_\phi$
(for example through the misalignment mechanism \cite{Preskill:1982cy,Abbott:1982af,Dine:1982ah}),
so this tuning may have an anthropic origin \cite{Freivogel:2008qc}. See also Ref.~\cite{Banerjee:2022sqg} for  possible mechanisms to obtain light scalar particles without fine-tuning.
On the other hand, a coupling of the form
\[
\De\scr{L} = -\frac{\la_\phi}{4!} \phi^4
\]
is allowed by all symmetries and is also UV sensitive.%
\footnote{We assume a $\mathbb{Z}_2$ symmetry under which $\phi \mapsto -\phi$,
so that terms with odd powers of $\phi$ are forbidden.}
This coupling is tightly constrained by constraints from structure formation
\cite{Arvanitaki:2014faa}
\[
\eql{lambdaphibound}
\la_\phi \lsim 3 \times 10^{-7} \left( \frac{m_\phi}{\eV} \right)^4.
\]
A value of $\la_\phi$ that violates this bound modifies the spectrum of
density fluctuations, but structure formation still takes place,
so there is no obvious anthropic constraint on $\la_\phi$.

The simplest UV completion of the model comes from adding the scalar $\phi$
to the Standard Model with the most general renormalizable couplings
compatible with the $\phi \mapsto -\phi$ symmetry:
\[
\De\scr{L}_\text{SM} =- \sfrac 12 m_\phi^2 \phi^2
- \frac{\la_\phi}{4!} \phi^4
- \sfrac 12 \ep \phi^2 H^\dagger H.
\]
This gives a $\phi$ dependent shift
in the Higgs VEV:
\[
\frac{\de v}{v} = \frac{\ep \phi^2}{2m_h^2},
\]
where $m_h = 125$~GeV is the physical Higgs mass.
Below the electroweak symmetry breaking scale, this induces
various couplings of $\phi^2$ to standard model fields.
For example, the coupling to SM fermions $\psi$ is given by
\[
\De\scr{L}_\text{eff} = -\sum_\psi \frac{1}{2f_\psi} \phi^2 \bar\psi \psi
\qquad
\frac{1}{f_\psi} = \frac{\ep \gap m_\psi}{m_h^2}.
\]
For light fermions such as the electron and the up and down quarks,
this is suppressed by the fermion mass.
The low-energy couplings that are not suppressed in this way are couplings to
the SM gauge field strength operators, for example
$\phi^2 G^{\mu\nu}_a G\sub{\mu\nu a}$, 
where $G_{\mu\nu a}$ is the gluon field strength.
Integrating out the heavy quarks gives a common contribution to the neutron and proton
masses
\[
\frac{1}{f_p} \simeq \frac{1}{f_n}
\simeq \frac{2}{9} \frac{\ep\gap m_p}{m_h^2}.
\]
At low energies, this model dominantly couples to the nucleons,
and this coupling is approximately isospin preserving.
This corresponds to the weaker nucleosynthesis constraint in the
left panel of Fig.~\ref{fig:constraints}.%
\footnote{The coupling to the electron can
suppress value of $\avg{\phi^2}$ during nucleosynthesis,
resulting in a weaker bound than if we neglect the electron
coupling \cite{Bouley:2022eer}.
This effect is not included in Fig.~\ref{fig:constraints},
so this bound is very conservative for this model.}


In this model, the $\phi^4$ coupling has a 
UV divergent 1-loop contribution of order
\[
\De \la_\phi \sim \frac{\ep^2}{16\pi^2} \ln\frac{\La_\text{UV}}{m_h},
\]
where $\La_\text{UV}$ is the UV cutoff, which we identify
with the scale of new physics.
Comparing this to the bound \Eq{lambdaphibound}
from structure formation we have
\[
\frac{\De \la_\phi}{\la_{\phi\gap\text{bound}}} 
\sim 10^{-6} \left( \frac{m_\phi}{\text{eV}} \right)^{-4}
\left( \frac{f}{10^{10}\GeV} \right)^{-2},
\]
where we have assumed that $\log(\La_\text{UV}/m_h) \sim 1$.
As long as this is ratio is smaller than 1, the model is
not fine-tuned.

Next, we consider another UV completion which can dominantly
couple to electrons at low energies.
This comes from adding the following interaction to the SM:
\[
\scr{L}_\text{SM} = -\frac{1}{2 M_e^2} \phi^2 (L H e^c + \hc).
\]
This requires UV completion at a scale of order $M_e$, so we require
$M_e \gsim \text{TeV}$.
Below the electroweak symmetry breaking scale, this generates a coupling
of $\phi^2$ to the electron given by
\[
\frac{1}{f_e} = \frac{m_e}{M_e^2}.
\]
In this UV completion, the dark matter couples dominantly to the
electron at low energies.

We now discuss the fine-tuning of the quartic coupling in this model.
The dominant UV divergent contribution is given by a 2-loop diagram,
and is of order

\[
\De\la_\phi \sim \left( \frac{1}{16\pi^2} \right)^2
\frac{\La_\text{UV}^4}{M^4}
\]
which gives
\[
\frac{\De \la_\phi}{\la_{\phi\gap\text{bound}}}
\sim \left( \frac{m_\phi}{\text{eV}} \right)^{-4}
\left( \frac{\La_\text{UV}}{\text{TeV}} \right)^4
\left( \frac{f}{10^{10}\GeV} \right)^{-2}.
\]

A variation of this model is to couple dominantly to a light
quark, for example the down quark:
\[
\scr{L}_\text{SM} = -\frac{1}{2 M_d^2} \phi^2 (Q H d^c + \hc).
\]
Below the electroweak symmetry breaking scale, the dominant
coupling is to the down quark, with
\[
\frac{1}{f_d} = \frac{m_d}{M_d^2}.
\]
This violates isospin maximally, corresponding to the 
stronger nucleosynthesis bound in the left panel of Fig.~\ref{fig:constraints}.

\bibliographystyle{utphys}
\bibliography{references}

\providecommand{\href}[2]{#2}\begingroup\raggedright\begin{thebibliography}{10}

\bibitem{Wagner:2012ui}
T.~A. Wagner, S.~Schlamminger, J.~H. Gundlach, and E.~G. Adelberger,
  ``{Torsion-balance tests of the weak equivalence principle},''
  \href{http://dx.doi.org/10.1088/0264-9381/29/18/184002}{{\em Class. Quant.
  Grav.} {\bfseries 29} (2012) 184002},
  \href{http://arxiv.org/abs/1207.2442}{{\ttfamily arXiv:1207.2442 [gr-qc]}}.

\bibitem{Schlamminger:2007ht}
S.~Schlamminger, K.~Y. Choi, T.~A. Wagner, J.~H. Gundlach, and E.~G.
  Adelberger, ``{Test of the equivalence principle using a rotating torsion
  balance},'' \href{http://dx.doi.org/10.1103/PhysRevLett.100.041101}{{\em
  Phys. Rev. Lett.} {\bfseries 100} (2008) 041101},
  \href{http://arxiv.org/abs/0712.0607}{{\ttfamily arXiv:0712.0607 [gr-qc]}}.

\bibitem{Lee:2020zjt}
J.~G. Lee, E.~G. Adelberger, T.~S. Cook, S.~M. Fleischer, and B.~R. Heckel,
  ``{New Test of the Gravitational $1/r^2$ Law at Separations down to 52
  $\mu$m},'' \href{http://dx.doi.org/10.1103/PhysRevLett.124.101101}{{\em Phys.
  Rev. Lett.} {\bfseries 124} no.~10, (2020) 101101},
  \href{http://arxiv.org/abs/2002.11761}{{\ttfamily arXiv:2002.11761
  [hep-ex]}}.

\bibitem{Hees:2018fpg}
A.~Hees, O.~Minazzoli, E.~Savalle, Y.~V. Stadnik, and P.~Wolf, ``{Violation of
  the equivalence principle from light scalar dark matter},''
  \href{http://dx.doi.org/10.1103/PhysRevD.98.064051}{{\em Phys. Rev. D}
  {\bfseries 98} no.~6, (2018) 064051},
  \href{http://arxiv.org/abs/1807.04512}{{\ttfamily arXiv:1807.04512 [gr-qc]}}.

\bibitem{Banerjee:2022sqg}
A.~Banerjee, G.~Perez, M.~Safronova, I.~Savoray, and A.~Shalit, ``{The
  phenomenology of quadratically coupled ultra light dark matter},''
  \href{http://dx.doi.org/10.1007/JHEP10(2023)042}{{\em JHEP} {\bfseries 10}
  (2023) 042}, \href{http://arxiv.org/abs/2211.05174}{{\ttfamily
  arXiv:2211.05174 [hep-ph]}}.

\bibitem{Fukuda:2018omk}
H.~Fukuda, S.~Matsumoto, and T.~T. Yanagida, ``{Direct Detection of Ultralight
  Dark Matter via Astronomical Ephemeris},''
  \href{http://dx.doi.org/10.1016/j.physletb.2018.12.038}{{\em Phys. Lett. B}
  {\bfseries 789} (2019) 220--227},
  \href{http://arxiv.org/abs/1801.02807}{{\ttfamily arXiv:1801.02807
  [hep-ph]}}.

\bibitem{Graham:2015ifn}
P.~W. Graham, D.~E. Kaplan, J.~Mardon, S.~Rajendran, and W.~A. Terrano, ``{Dark
  Matter Direct Detection with Accelerometers},''
  \href{http://dx.doi.org/10.1103/PhysRevD.93.075029}{{\em Phys. Rev. D}
  {\bfseries 93} no.~7, (2016) 075029},
  \href{http://arxiv.org/abs/1512.06165}{{\ttfamily arXiv:1512.06165
  [hep-ph]}}.

\bibitem{Carney:2019cio}
D.~Carney, A.~Hook, Z.~Liu, J.~M. Taylor, and Y.~Zhao, ``{Ultralight dark
  matter detection with mechanical quantum sensors},''
  \href{http://dx.doi.org/10.1088/1367-2630/abd9e7}{{\em New J. Phys.}
  {\bfseries 23} no.~2, (2021) 023041},
  \href{http://arxiv.org/abs/1908.04797}{{\ttfamily arXiv:1908.04797
  [hep-ph]}}.

\bibitem{Buscaino:2015fya}
B.~Buscaino, D.~DeBra, P.~W. Graham, G.~Gratta, and T.~D. Wiser, ``{Testing
  long-distance modifications of gravity to 100 astronomical units},''
  \href{http://dx.doi.org/10.1103/PhysRevD.92.104048}{{\em Phys. Rev. D}
  {\bfseries 92} no.~10, (2015) 104048},
  \href{http://arxiv.org/abs/1508.06273}{{\ttfamily arXiv:1508.06273 [gr-qc]}}.

\bibitem{Raffelt:1990yz}
G.~G. Raffelt, ``{Astrophysical methods to constrain axions and other novel
  particle phenomena},''
  \href{http://dx.doi.org/10.1016/0370-1573(90)90054-6}{{\em Phys. Rept.}
  {\bfseries 198} (1990) 1--113}.

\bibitem{Olive:2007aj}
K.~A. Olive and M.~Pospelov, ``{Environmental dependence of masses and coupling
  constants},'' \href{http://dx.doi.org/10.1103/PhysRevD.77.043524}{{\em Phys.
  Rev. D} {\bfseries 77} (2008) 043524},
  \href{http://arxiv.org/abs/0709.3825}{{\ttfamily arXiv:0709.3825 [hep-ph]}}.

\bibitem{DeRocco:2020xdt}
W.~DeRocco, P.~W. Graham, and S.~Rajendran, ``{Exploring the robustness of
  stellar cooling constraints on light particles},''
  \href{http://dx.doi.org/10.1103/PhysRevD.102.075015}{{\em Phys. Rev. D}
  {\bfseries 102} no.~7, (2020) 075015},
  \href{http://arxiv.org/abs/2006.15112}{{\ttfamily arXiv:2006.15112
  [hep-ph]}}.

\bibitem{Stadnik:2015kia}
Y.~V. Stadnik and V.~V. Flambaum, ``{Can dark matter induce cosmological
  evolution of the fundamental constants of Nature?},''
  \href{http://dx.doi.org/10.1103/PhysRevLett.115.201301}{{\em Phys. Rev.
  Lett.} {\bfseries 115} no.~20, (2015) 201301},
  \href{http://arxiv.org/abs/1503.08540}{{\ttfamily arXiv:1503.08540
  [astro-ph.CO]}}.

\bibitem{Sibiryakov:2020eir}
S.~Sibiryakov, P.~S\o{}rensen, and T.-T. Yu, ``{BBN constraints on
  universally-coupled ultralight scalar dark matter},''
  \href{http://dx.doi.org/10.1007/JHEP12(2020)075}{{\em JHEP} {\bfseries 12}
  (2020) 075}, \href{http://arxiv.org/abs/2006.04820}{{\ttfamily
  arXiv:2006.04820 [hep-ph]}}.

\bibitem{Bouley:2022eer}
T.~Bouley, P.~S\o{}rensen, and T.-T. Yu, ``{Constraints on ultralight scalar
  dark matter with quadratic couplings},''
  \href{http://dx.doi.org/10.1007/JHEP03(2023)104}{{\em JHEP} {\bfseries 03}
  (2023) 104}, \href{http://arxiv.org/abs/2211.09826}{{\ttfamily
  arXiv:2211.09826 [hep-ph]}}.

\bibitem{MICROSCOPE:2022doy}
{\bfseries MICROSCOPE} Collaboration, P.~Touboul {\em et~al.}, ``{MICROSCOPE
  Mission: Final Results of the Test of the Equivalence Principle},''
  \href{http://dx.doi.org/10.1103/PhysRevLett.129.121102}{{\em Phys. Rev.
  Lett.} {\bfseries 129} no.~12, (2022) 121102},
  \href{http://arxiv.org/abs/2209.15487}{{\ttfamily arXiv:2209.15487 [gr-qc]}}.

\bibitem{Nobili:2012uj}
A.~M. Nobili {\em et~al.}, ``{'Galileo Galilei' (GG): Space test of the weak
  equivalence principle to 10(-17) and laboratory demonstrations},''
  \href{http://dx.doi.org/10.1088/0264-9381/29/18/184011}{{\em Class. Quant.
  Grav.} {\bfseries 29} (2012) 184011}.

\bibitem{VanTilburg:2024tst}
K.~Van~Tilburg, ``{Wake Forces},''
  \href{http://arxiv.org/abs/2401.08745}{{\ttfamily arXiv:2401.08745
  [hep-ph]}}.

\bibitem{Landau:1991wop}
L.~D. Landau and E.~M. Lifshits, {\em {Quantum Mechanics}: {Non-Relativistic
  Theory}}, vol.~v.3 of {\em Course of Theoretical Physics}.
\newblock Butterworth-Heinemann, Oxford, 1991.

\bibitem{Preskill:1982cy}
J.~Preskill, M.~B. Wise, and F.~Wilczek, ``{Cosmology of the Invisible
  Axion},'' \href{http://dx.doi.org/10.1016/0370-2693(83)90637-8}{{\em Phys.
  Lett. B} {\bfseries 120} (1983) 127--132}.

\bibitem{Abbott:1982af}
L.~F. Abbott and P.~Sikivie, ``{A Cosmological Bound on the Invisible Axion},''
  \href{http://dx.doi.org/10.1016/0370-2693(83)90638-X}{{\em Phys. Lett. B}
  {\bfseries 120} (1983) 133--136}.

\bibitem{Dine:1982ah}
M.~Dine and W.~Fischler, ``{The Not So Harmless Axion},''
  \href{http://dx.doi.org/10.1016/0370-2693(83)90639-1}{{\em Phys. Lett. B}
  {\bfseries 120} (1983) 137--141}.

\bibitem{Freivogel:2008qc}
B.~Freivogel, ``{Anthropic Explanation of the Dark Matter Abundance},''
  \href{http://dx.doi.org/10.1088/1475-7516/2010/03/021}{{\em JCAP} {\bfseries
  03} (2010) 021}, \href{http://arxiv.org/abs/0810.0703}{{\ttfamily
  arXiv:0810.0703 [hep-th]}}.

\bibitem{Arvanitaki:2014faa}
A.~Arvanitaki, J.~Huang, and K.~Van~Tilburg, ``{Searching for dilaton dark
  matter with atomic clocks},''
  \href{http://dx.doi.org/10.1103/PhysRevD.91.015015}{{\em Phys. Rev. D}
  {\bfseries 91} no.~1, (2015) 015015},
  \href{http://arxiv.org/abs/1405.2925}{{\ttfamily arXiv:1405.2925 [hep-ph]}}.

\end{thebibliography}\endgroup

\end{document}